# Cavitation and charge separation in laser-produced copper and carbon plasma in transverse magnetic field

Narayan Behera*, Ajai Kumar and R. K. Singh

Institute for Plasma Research, Bhat, Gandhinagar - 382 428, India

*Email: nbehera@ipr.res.in

**Abstract**

In the present work, we report the dynamics and geometrical features of the plasma plume formed by the laser ablation of copper and graphite (carbon) targets in the presence of different transverse magnetic field. This work emphasizes on the effect of atomic mass of the plume species on the diamagnetic behaviour and geometrical aspect of the expanding plasma plume in the magnetic field. The time-resolved analysis of the simultaneously captured two directional images in orthogonal to the expansion axis is carried out for the comparative study of projected three-dimensional structure of copper and carbon plasma plume. In the presence of magnetic field, sharp differences are observed between the copper and carbon plasma plumes in terms of formation of diamagnetic cavity and structure formation. An elliptical cavity-like structure is observed in case of copper plasma plume which attains the sharp conical shape with increasing the time delay or magnetic field strength. On the other hand, splitted carbon plasma plume appears as a Y-shape structure in the presence of magnetic field where the cavity-like structure is not observed for the considered time and magnetic field. Based on the modified energy balance relation for the elliptic cylindrical geometry, we have also simulated the dynamics of the plume which is in close agreement with observed plasma expansion in diamagnetic and non-diamagnetic regions.

# 1. Introduction

The study of laser-produced plasma in the presence of external magnetic field continues to attract the attention because of its applications in several area like tokamak plasma, space plasmas such as the study of artificial comets, propulsion of space vehicles using laser ablation, investigation of acceleration of stellar winds, laboratory plasmas such as increase in the detection sensitivity of laser-induced breakdown spectroscopy [1-9], manipulation of plasma plume in pulsed laser deposition [10-12], debris mitigation [13] etc. Apart from the applied research, the plasma plume-magnetic field interaction is useful to understand several fundamental phenomena like conversion of kinetic energy into plasma thermal energy, plume confinement, ion acceleration/deceleration, emission enhancement/decrease and plasma instabilities [1-8]. The plasma parameters like electron temperature, electron density, Larmor motion of electrons/ions and applied magnetic field strength are the key factors to govern the dynamical and geometrical behaviour of plasma plume in the presence of the magnetic field. The diamagnetism of the plasma plume has been extensively analyzed in terms of $J \times B$ force interaction with the assumption plasma plume having spherical symmetry in most of the reported works [2,14,15]. Several pioneer work have been done related to plasma plume-magnetic field interactions in uniform and non-uniform magnetic field [4,14,15], formation of diamagnetic cavity associated with flute-like structures [16,17], plasma oscillations [18], edge instability and dramatic structuring of the plume [19, 20] and sub-Alfvenic plasma expansion [21]. However at later stage of expansion, diffusion of magnetic field into the plasma plume and subsequent polarization and $E \times B$ drift of the plume across the field is largely ignored in the reported work especially in case of moderate magnetic field (< 1 T) and low laser energy of order of few hundred mJ.

Further shape, size and geometrical structure of plasma plume in the presence of magnetic is another important aspect in terms of its utilization in practical applications [10-12].Various time-resolved 2D fast imaging have been carried out which gives the general features of the geometrical aspect and dynamics of the expanding plasma plume in transverse magnetic field produced by the pair of permanent bar magnets [2,4,15]. However due to the experimental constraint (plasma imaging is not possible along the magnetic field direction), asymmetric structure formation is not captured in the majority of reported works [2,4,15]. Recently we have developed an experimental setup based on Helmholtz coil and pulse power system [22] which is capable of capturing two-directional imaging in orthogonal to the expansion axis (across as well as along the magnetic field lines). This gives complete information about the asymmetric structure of the plasma plume. Further, formation and collapse of the diamagnetic cavity and its shape and size in the expanding plasma plume basically depends on the Larmor motion of electrons and ions. Therefore, it is of interest to carry out a comparative study of lighter and heavier plume species in different magnetic field with regards to their expansion in diamagnetic and non-diamagnetic regions and structure formation.

Based on the above considerations, we have carried out a systematic experimental study of the effect of different magnetic field on the dynamical and geometrical aspect of the expanding copper and carbon plasma plume. In this context, the simultaneously captured images on orthogonal directions to the expansion direction are analysed as function of time in different magnetic field varying from 0.13 to 0.5 T. In this report, emphasis is given to the comparison of the dynamics, geometrical structure and diamagnetism of the carbon and copper plasma plume in identical experimental conditions. The observed results are validated from the numerical simulation based on a theoretical model. Isotropic graphite and electrolytic tough pitch (ETP) copper are chosen as a target because of the large difference in their atomic masses and also because of their applications in thermonuclear devices, semiconductor and wide field of industries. Further isotropic graphite is favourable material as a target in laser ion source [23].

## 2. Experimental scheme

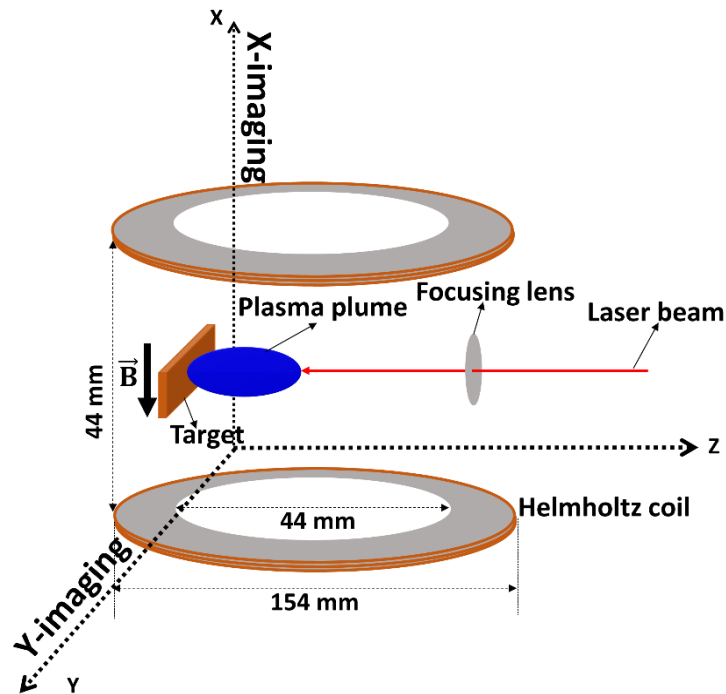

**Fig. 1.** Schematic of the experimental setup.

Detailed description of the experimental setup has been reported in elsewhere [22,24]. A brief summary which is important for the present report is summarized here. The plasma plume is created in a vacuum chamber having a base pressure less than $10^{-6}$ Torr using an 1064 nm, 8 ns FWHM pulsed Nd:YAG laser. The pulse energy and spot diameter of the laser beam are set to 100 mJ and 0.5 mm respectively. The different transverse magnetic field ranging from 0.13 to 0.57 T is produced by a Helmholtz coil along with the indigenously developed pulse power system. The width of the magnetic field profile is ~ 1 ms having flat-top region ~ 20 μs. Since the plasma duration (few microseconds) is less than the flat-top of the magnetic field profile and plasma formation is synchronized with the flat-top region, therefore it is expected that plasma expands in the uniform magnetic field. In contrast to commonly used permanent bar magnets, the Helmholtz coil facilitates to capture simultaneous images of expanding plasma plume along as well as across the magnetic field lines. For that two mutually synchronized ICCD cameras placed on orthogonal directions to the expansion axis as shown in **Fig. 1**. The copper and graphite plates are mounted on a movables target holder through a vacuum compatible feed-through and placed in between the coils. The temporal evolution of the plasma plume has been captured by varying the time delay from 100 to 1200 ns between the laser pulse and the opening time of the ICCD gate. Gate width is set as 5-25 ns. A micro-controller based timing controller is used to synchronise the ICCD cameras and magnetic field with laser pulse. The dimensions of the plume are estimated by the image of known dimension mesh and digital image processing technique using MATLAB algorithm.

## 3. Results and discussion

Effect of magnetic field on the plume species of different atomic masses has been studied using copper and graphite as target materials. The dynamics and structural behaviour of the plasmas have been compared by analysing the two-directional imaging in orthogonal to the expansion axis (along and across the magnetic field lines). The different transverse magnetic fields are produced by Helmholtz coil coupled with the pulsed power system.

### 3.1. Expansion in the absence of magnetic field

The sequence of time-resolved images of the electronically excited plume species of copper and carbon plasma are captured orthogonal to the expansion axis (z-axis) that is along the x-axis (referred as X-imaging) and y-axis (referred as Y-imaging) in the absence of magnetic field as shown in **Fig. 2**. The time delay with respect to laser pulse is varying from 200 to 1200 ns. The background pressure is set as $10^{-6}$ Torr. Each image represents the spectrally integrated emission intensity in the range 350-750 nm, emitted from plume species.

In the absence of the external magnetic field, the expansion of the plasma plume is mainly governed by the initial pressure gradient inside the plume and it is well treated as adiabatic expansion [25]. In the case of copper, the expanding plasma plume takes nearly an ellipsoidal shape due to the large pressure gradient in the direction perpendicular to the target surface. Both X-imaging and Y-imaging of the plasma plumes are nearly identical as shown in **Fig. 2a** which established the ellipsoidal structure of the expanding plume. The slight differences in the intensities of X-imaging and Y-imaging might be related to the sensitivity of two cameras. In contrast to the copper plasma, carbon plasma plume is splitted into two well-separated components, the leading component and slow moving component closer to the target plate which are designated as fast and slow components as shown in **Fig. 2b**. The luminosity of the plasma plume gradually decreases with time because of rapid decrease of electron temperature and density and hence the probability of excitation of plume species. After 1200 ns delay time the emission intensity is beyond the detection limit of the ICCD.

Linear expansion of the copper plasma in the absence of magnetic field can be seen on the plume front position vs. time plot as shown in **Fig. 3**. Here the estimated error in plume length is found to be <3% which falls within the size of symbol. The average expansion velocity is obtained by slope of the linear fit of the data which is $4.05 \times 10^4$ m/s. The position of both fast and slow components of carbon plasma with time is also included in **Fig. 3**. Slopes of the linear fit curve give the velocity of fast and slow component of the carbon plasma as $7.0 \times 10^4$ m/s and $1.2 \times 10^4$ m/s respectively.

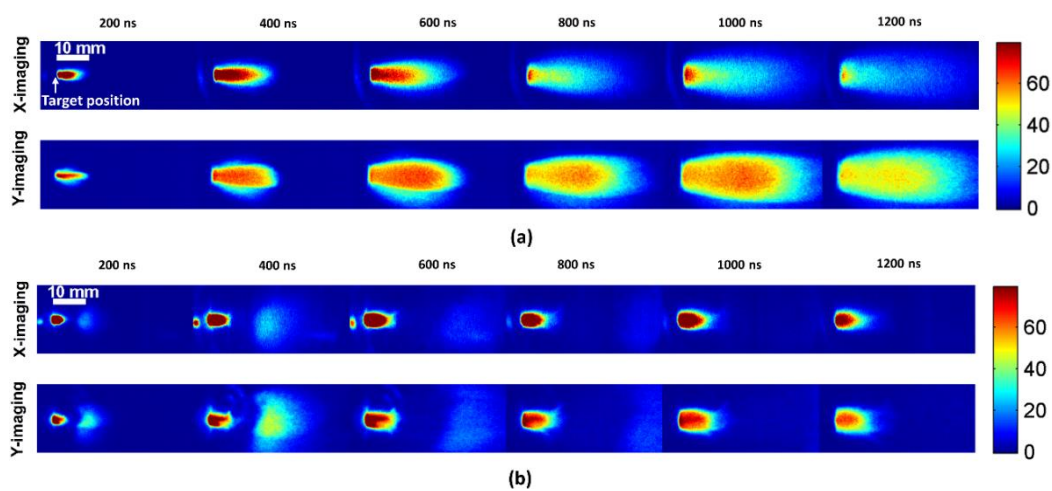

**Fig. 2.** The sequence of images of the expanding (a) copper and (b) carbon plasma plume in the absence of magnetic field at different delay times.

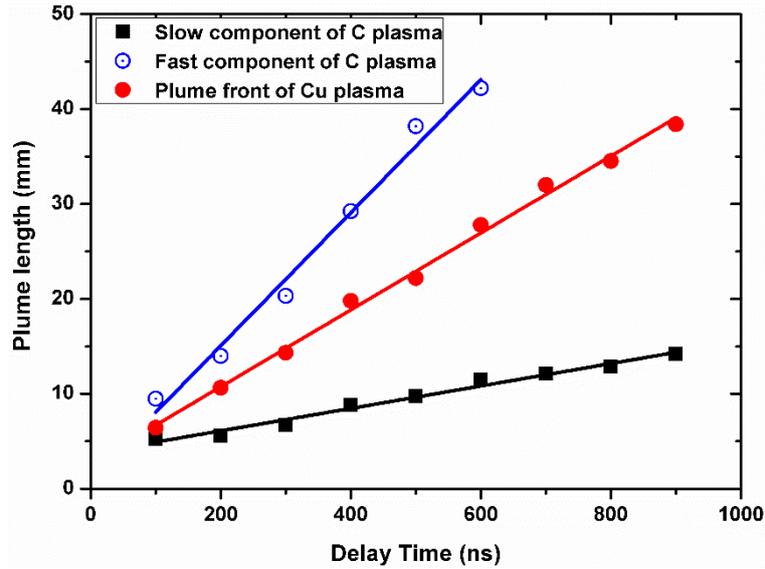

**Fig. 3**. Variation of length of plume front of copper plasma and slow and fast components of carbon plasma as a function of delay time in the absence of magnetic field. Solid lines represent the linear fit of the data.

In order to understand the plume splitting in vacuum, especially in low atomic mass (for example, carbon) we have analysed the images of the plasma plume formed by copper, aluminium and carbon in identical experimental conditions without external magnetic field. Here the image of aluminium plasma [22] is introduced to see the effect of mass of ablated species on the splitting pattern of the plasma. The images of the copper, aluminium and carbon plasma along x-axis (X-imaging) in vacuum and at 400 ns delay time is shown in **Fig. 4**. As already mentioned, copper plasma has a well-defined ellipsoidal shape where the signature of plume splitting is not observed. In aluminium having atomic mass roughly half of the copper, the plume splitting as two lobes is clearly visible but it is not separated apart as observed in carbon plasma. This means that plume splitting becomes apparent with decrease of atomic mass of the targets where the separation between fast and slow components is increasing in lighter plume species. This could be understood with double layer formation in laser-produced plasma [26].The energetic electrons of the plasma try to escape from plasma due to its high velocity and hence induce the electric field [27,28]. The ions are accelerated by this electric field forming the leading edge of the plasma (fast component) and leaving behind a neutral dominant component (slow component). As carbon ion is lighter, it accelerates with higher velocity in comparison to aluminium and copper and hence well separated from its neutral dominated component.

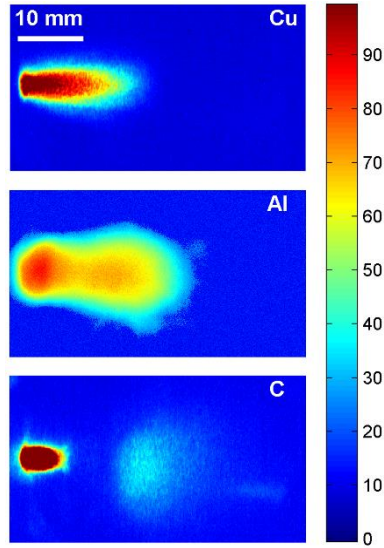

**Fig. 4.** Images of the expanding copper, aluminium and carbon plasma in the absence of magnetic field at 400 ns delay time.

**3.2. Expansion in the presence of 0.13 T magnetic field**

The dynamics of structure formation of the plasma plume in presence of magnetic field are studied in identical experimental conditions as infield-free case. The sequence of images of the expanding copper and carbon plasma plume at 0.13 T transverse magnetic field at different delay time varying from 200 ns to 1200 ns are shown in **Fig. 5**. With 0.13 T applied magnetic field, laser-produced copper plasma shows similar behaviour like other metallic plasma such as aluminium [22] as shown in **Fig. 5a**. We have observed an increase of emission intensity as compared to the field-free case. The diamagnetic cavity is observed as shown in X-imaging of **Fig. 5a** up to the considered time. Three-dimensional shape of the diamagnetic cavity is an elliptic cylinder which is confirmed from projected images with the combination of X and Y-imaging.

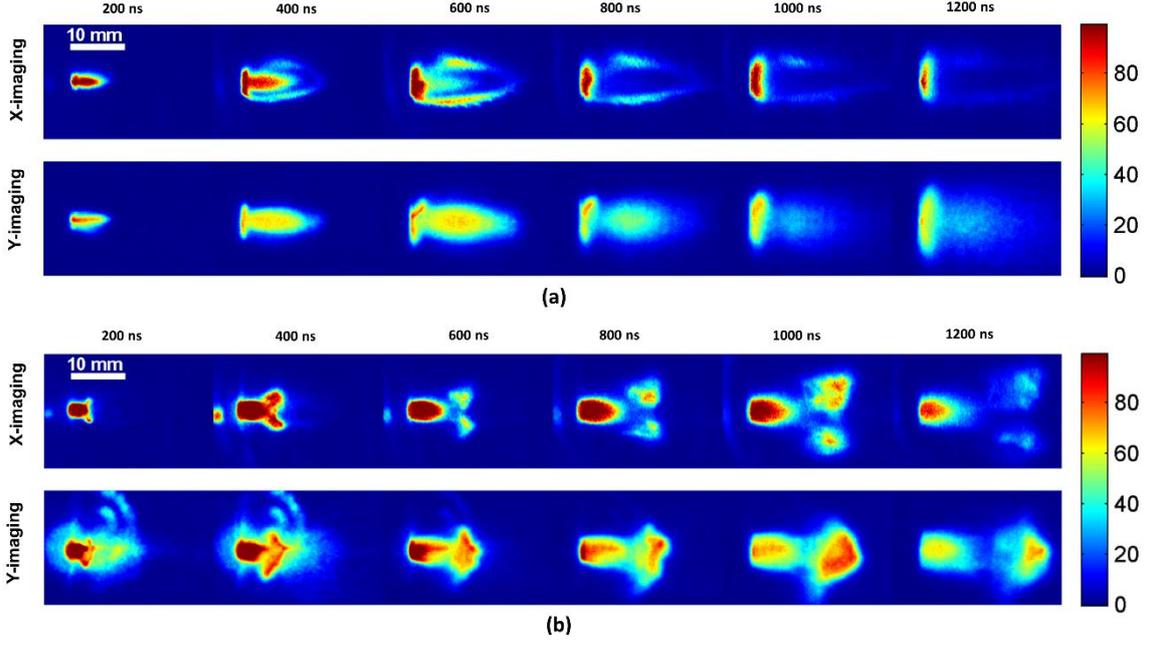

**Fig. 5.** The sequence of images of the expanding (a) copper and (b) carbon plasma plume at 0.13 T transverse magnetic field at different delay times.

The formation of diamagnetic cavity in copper plasma could be explained by the model proposed by Winske [17]. The estimated Larmor radius of electrons and Cu ion ions in presence of 0.13 T magnetic field are 1.77 μm and 20.5 cm respectively. Here the plasma parameters that is electron temperature ~ 1.0 eV and electron density ~ $1.52 \times 10^{16}$ cm$^{-3}$ are taken from the reported result [1] in similar experimental conditions. Since ion Larmor radius is larger than the observed dimensions of the plume therefore ions are treated as non-magnetized whereas electrons are magnetized [17]. In this scenario, the electrons are confined in small volume and ions expand radially outward and form a cavity-like structure because of an induced radial electric field **E** pointing inward direction. The magnetized electrons experienced an azimuthal drift (= (E×B)/B$^2$) relative to ions. This azimuthal drift induces the diamagnetic current and magnetic field in opposite to applied magnetic field direction. The diamagnetic cavity sustains as long as diamagnetic current/azimuthal drift exists.

Further, the plasma beta (thermal beta, $\beta_t$), defined as the ratio between plasma thermal pressure ($n_eT_e$) to the magnetic pressure ($B^2/2\mu_0$), plays an important role to govern the dynamics of plasma plume in the magnetic field. The plasma plume decelerates and stops when plasma thermal pressure is balanced by magnetic pressure (i.e., $\beta_t = 1$). However, the laser plasma expands beyond the $\beta_t = 1$ because of large directed energy ($m_i n_e v_0^2/2$) along the expansion axis. Therefore, total beta ($\beta_{total} = (m_i n_e v_0^2/2 + n_e T_e)/(B^2/2\mu_0)$) is important in the present case which is always greater than one for the considered time. Here $n_e$ is the electron density and $T_e$ is the electron temperature, $m_i$ is the mass of ion, B is the applied magnetic field, $\mu_0$ is the magnetic permeability and $v_0$ is the expansion velocity of the plasma plume.

In order to understand the plasma plume-magnetic field interactions, some parameters related to diamagnetism of the plasma plume, for example, cavity size, stopping distance ($R_B$) and expansion duration ($t_c$) are estimated by conservation of the energy balance equation. The $t_c$ is the characteristic time where axial plasma dimension is equal to bubble radius $R_B$ and plume tending to stagnation. The energy balance equation for the plasma expansion across the magnetic field [21,29,30] is written as,

$$E = \frac{1}{2}Mv^2 + \left(\frac{\pi abh}{4}\right)\left(\frac{B^2}{2\mu_0}\right) = \frac{1}{2}Mv^2 + \left(\frac{\pi a^3}{4c_1 c_2}\right)\left(\frac{B^2}{2\mu_0}\right) \ldots \ldots \ldots \ldots (1)$$

Here M is the mass of the plume, v is the expansion velocity and B is the magnetic field strength and a, b and h are the major axis, minor axis and height of the elliptic cylindrical structure of the plasma plume respectively. For simplicity, we have defined $c_1 = a/b$ and $c_2 = a/h$ which are estimated from the recorded image as ~ 2. In ideal condition, we have approximated the initial condition as, $a(0) = 0$ and $v(0) = v_0$ (initial expansion velocity of plasma plume) and therefore, $E = ½ Mv_0^2 = E_{lpp}$. Here, the kinetic energy of the laser plasma, $E_{lpp}$ is approximated as half of the laser beam energy [19,21,31]. Further for the boundary condition ($a(t_c) = R_B$, $v(t_c) = 0$),

$$R_B = \left(\frac{8\mu_0 c_1 c_2 E_{lpp}}{\pi B^2}\right)^{\frac{1}{3}} \ldots \ldots \ldots \ldots (2)$$

Thus, the equation of motion of plasma plume in magnetic field is deduce from eq(1) as

$$\frac{d^2 a}{dt^2} + \left(\frac{3}{2}\frac{v_0^2}{R_B^3}\right)a^2 = 0 \ldots \ldots \ldots \ldots (3)$$

Here, decelerating force experienced by the plasma plume due to magnetic pressure is

$$g_{eff} = \frac{3}{2}\frac{v_0^2}{R_B^3}a^2 \ldots \ldots \ldots \ldots (4)$$

Eq(3) is a second order homogeneous nonlinear ordinary differential equation. We can solve this equation to find *a*(t) and v(t). The expression of velocity, time and expansion duration are obtained as

$$v = v_0 \sqrt{1 - \left(\frac{a}{R_B}\right)^3} \quad \ldots \ldots \ldots \ldots \ldots (5)$$

$$t = \frac{1}{v_0} \int_0^a \frac{da}{\sqrt{1 - \left(\frac{a}{R_B}\right)^3}} = \left(\frac{a}{v_0}\right) \cdot {}_2F_1\left(\frac{1}{3}, \frac{1}{2}; \frac{4}{3}; \left(\frac{a}{R_B}\right)^3\right) \quad \ldots \ldots \ldots \ldots \ldots (6)$$

where ${}_2F_1(a, b; c; z)$ is the hypergeometric function.

$$t_c = \frac{1}{v_0} \int_0^{R_B} \frac{da}{\sqrt{1 - \left(\frac{a}{R_B}\right)^3}} \approx 1.4 \frac{R_B}{v_0} \quad \ldots \ldots \ldots \ldots \ldots (7)$$

Thus for the present case $E_{lpp}$ = 50 mJ, estimated $R_B$ is 33.58 mm and $t_c$ is 1160.83 ns for the copper plasma plume at 0.13 T magnetic field.

In order to demonstrate the plume expansion in the diamagnetic region, we have plotted plume front position vs. time plot in **Fig. 6**. It has been observed that the plume front position vs. time curve is flattened (that is approaching to stagnation)) ~ 1000 ns where the plume dimension is ~ 28 mm. This is close to the estimated $t_c$ and $R_B$. The expansion in the diamagnetic region is further validated with simulation. Since *a* can't express as an explicit function of t and therefore theoretical variation of plume dimension as function of time delay are obtained numerically by solving eq(3) using ode45 solver based on the Runge-Kutta method [32] with initial conditions, $a(0) = 0$ and $v(0) = v_0$. The simulation plot is also included in **Fig. 6**. Close agreement between the experimental data and simulation is observed in the diamagnetic region.

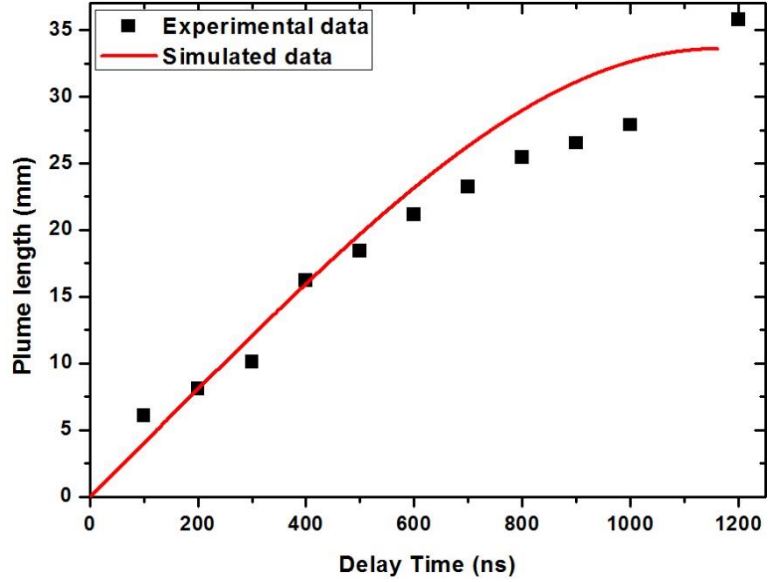

**Fig. 6.** Variations of plume length in expansion direction as a function of time delay in the presence of 0.13 T magnetic field. Solid line represents the simulated data.

The scenario is completely different in the presence of magnetic field in carbon plasma plume in comparison to copper plasma. The time-resolved images of carbon plasma plume in both X and Y-directions (X-imaging and Y-imaging) are shown in **Fig. 5b**. Unlike the cavity-like structure in copper plasma, there is no diamagnetic cavity observed in carbon plasma. The difference in atomic mass and charge state distributions in carbon and copper plasma might be responsible for the above observation. Here it should be noted that the difference in binding energy between the isotropic and amorphous graphite may affect the plasma compositions and hence present observations, however this is not studied in the present report. A well-separated neutral dominated slow component and ion dominated leading component of carbon plasma responded differently to the applied transverse magnetic field. The neutral dominant component doesn't show any significant change in presence of field. Interestingly, the ionic component bifurcated into two components perpendicular to the magnetic field direction as shown in X-imaging of **Fig. 5b**. The symmetrical bifurcation of the leading component of carbon plasma in the XZ-plane is confirmed by Y-imaging as shown in **Fig. 5b**. Thus the projected images of X-imaging and Y-imaging suggest the Y-shape structure formation of carbon plasma plume in presence of 0.13 T magnetic field. The absence of diamagnetic cavity and bifurcated structure of leading edge of the carbon plume infer that the magnetic field is not displaced by induced diamagnetic current rather it diffuses into the carbon plasma plume. This observation could be understood as follows. As already mentioned that neutral species are dominated in the slow component of the plasma plume and hence it remains unchanged in the presence of field as in the case of field-free shown in **Fig. 5b**. On the other hand in the charge particle dominated fast region, the electrons and ions are gyrated in opposite directions in the presence of a uniform magnetic field. It has been already discussed that electrons are confined within a small volume whereas ions expand radially outwards with its velocity. Due to the higher velocity of carbon ions, it may overcome the restoring force of confined electrons. In this scenario, it seems to be the radial electric field and hence diamagnetic current is not induced for the considered magnetic field strength. Basically, the azimuthal drift velocity that responsible for the diamagnetic current is regulated by the balance between driving (Larmor gyration) and restoring forces [33]. This might be the reason for the absence of cavity-like structure in carbon plasma plume. Therefore in the case of carbon, external magnetic field is diffused into the plasma plume and polarises the plume front in two components because of the Lorentz force (v×B).

**3.3. Plume expansion in variable magnetic field**

To illustrate the effect of magnetic field strength on the induced cavitation in the plasma plume, both X and Y-images of the plasma plume are recorded in different magnetic field varying from 0.13 to 0.5 T at onset of time delay 800 ns as shown in **Fig. 7**. The strong dependence of shape and size of the diamagnetic cavity on the magnetic field strength is clearly visible in X-imaging of copper plasma (**Fig. 7a**). With increase of magnetic field, the cavity-like structure appears compressed along the lateral direction, that is the minor axis of the elliptical cavity is decreased with increase of magnetic field and finally collapsed and formed a cone-like structure. On the other hand, plasma instability appears as striation-like structures at the surface of the plume are observed in Y-imaging of the copper plasma which are more pronounced at high magnetic field. The structured intensity pattern may be initiated due to velocity shear-driven instabilities [8,22].Interestingly, increase of the magnetic field strength from 0.13 to 0.5 T does not bring significant changes in the case of carbon plasma plume as depicted in **Fig. 7b**. Lateral separation between two lobes are slightly reduced with increase of magnetic field.

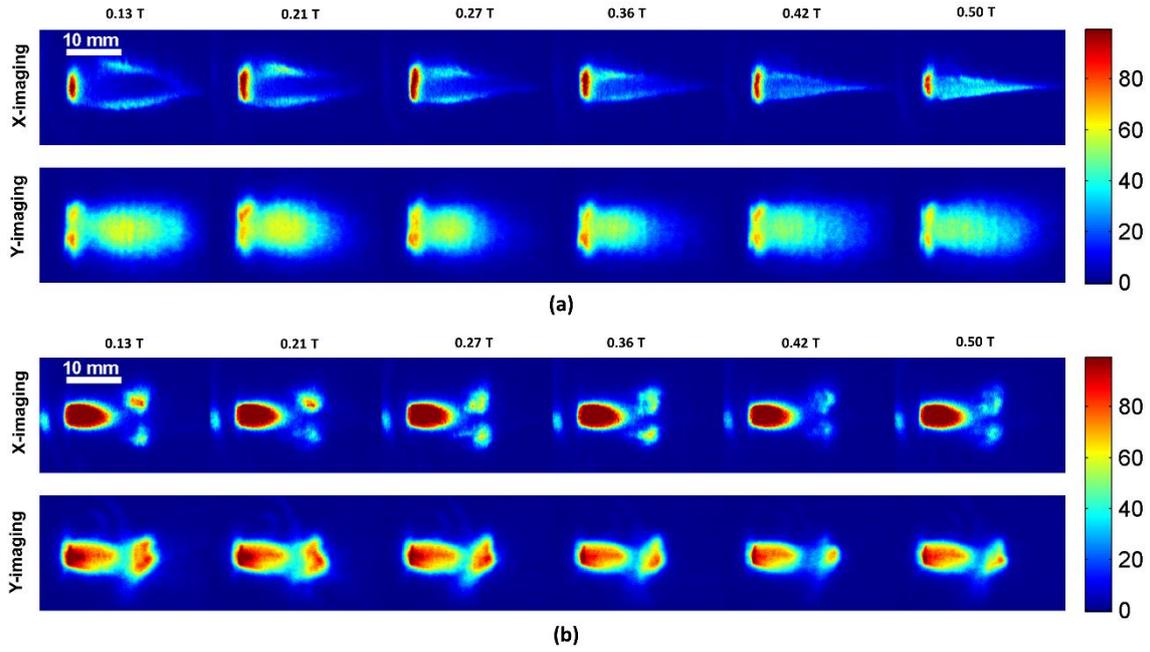

**Fig. 7.** Sequence of X and Y-images of copper (a) and carbon (b) plasma plume in different magnetic field, varying from 0.13 to 0.5 T. Time delay is set as 800 ns.

In order to show the plasma expansion in the non-diamagnetic limit where the cavitation in the plume is collapsed and magnetic field diffused into the plume, images of the expanding copper and carbon plasma plume at 0.5 T transverse magnetic field at different delay times are depicted in **Fig. 8**. The plasma plume shape and dynamics is significantly different in high magnetic field, B = 0.5 T as compared to expansion in low magnetic field, especially in case of copper plasma. In contrast to the low magnetic field (0.13 T), elliptic diamagnetic cavity is not observed in copper plasma plume at 0.5 T magnetic field for the considered delay time as shown in **Fig. 8a**. X-imaging in **Fig. 8a** clearly shows that copper plasma plume attains the cone-like structure with time and this conical structure becomes sharper with increasing the time delay. On the other hand, in case of Y-imaging, the plume dimension along the magnetic field direction (that is in the XZ-plane) increases with time. Striation-like structures are also observed in Y-imaging which is apparent at comparatively higher time delay.

The observed differences in the shape and dynamics of copper plasma plume in low and high magnetic field is because of the duration of diamagnetic region followed by diffusion of the magnetic field into the plume. At higher magnetic field, the diamagnetic nature of the plume collapses and the magnetic field is diffused in the plasma plume at an earlier time as compared to the low magnetic field. The magnetic diffusion time is estimated by the relation $t_d = \mu_0 \sigma R_B^2$. Here σ is the plasma conductivity $\frac{50\pi^{1/2}\varepsilon_0^2 T_e^{3/2}}{m_e^{1/2} e^2 Z \ln\Lambda}$, symbols have their standard meanings [14]. The estimated $t_d$ is found to be ~ 429.90 and 214.95 ns for the charge state Z = 1 and 2 respectively. We have also estimated the characteristic time ($t_c$) that is the maximum duration of the diamagnetic cavity using the energy balance equation which is ~ 472.88 ns. It has been observed that the copper plasma plume didn't experience the decelerating magnetic pressure and expanded linearly across 0.5 T magnetic field throughout the considered time range. This could be clearly visible in the plume front vs. time delay plot which is shown in **Fig. 9**. To compare the plume expansion in diamagnetic and non-diamagnetic regime, experimental data for 0.21 T along with the simulation plot using energy balance equation are also included in **Fig. 9**. Decelerating magnetic pressure up to the estimated characteristic time $t_c$ ~ 843.18 ns (estimated by equation (7)) and thereafter linear like expansion is clearly observed in case of 0.21 T magnetic field (**Fig. 9a**). On the other hand, in the case of 0.5 T field, the linear fit of the experimental data represents the free expansion of the plasma plume across the magnetic field (**Fig. 9b**). In this case a little discrepancy between the simulation and experiment is observed. Simulation predicts the diamagnetic region up 472.88 ns at 0.5 T field which is not observed in present observation.

The free expansion of the plasma plume across the magnetic field is attributed as diffuse of magnetic field into the plasma causes the polarization of ions and electrons in opposite direction due to Lorentz force (v×B).The charge separation in the plasma plume induces the electric field E and hence plasma expand with drift velocity (E × B)/B$^2$. In line with the above argument, the sharp conical structure of the plume could be understood with the strength of the induced electric field across the plume. The induced electric field E gradually decreases away from the plume centre because of the shielding effect. As a result, the particles on the surface of the polarized plasma experienced a comparative lesser electric field and hence drifted with less velocity in comparison to the inner layer [15,22,34,35]. Therefore, the inner layers of the plasma move ahead with respect to the outer layer and are repolarized. Thus, the successive lagging of the outer resulting a cone-like structure and it becomes more sharper with time as observed in **Fig. 8a**.

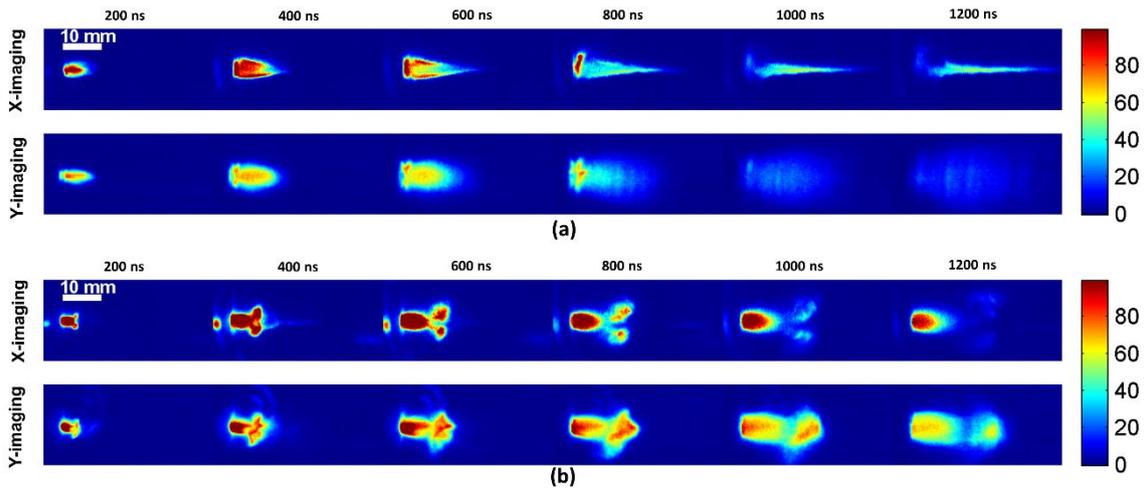

**Fig. 8.** The sequence of images of the expanding (a) copper and (b) carbon plasma plume at 0.5 T transverse magnetic field at different delay times.

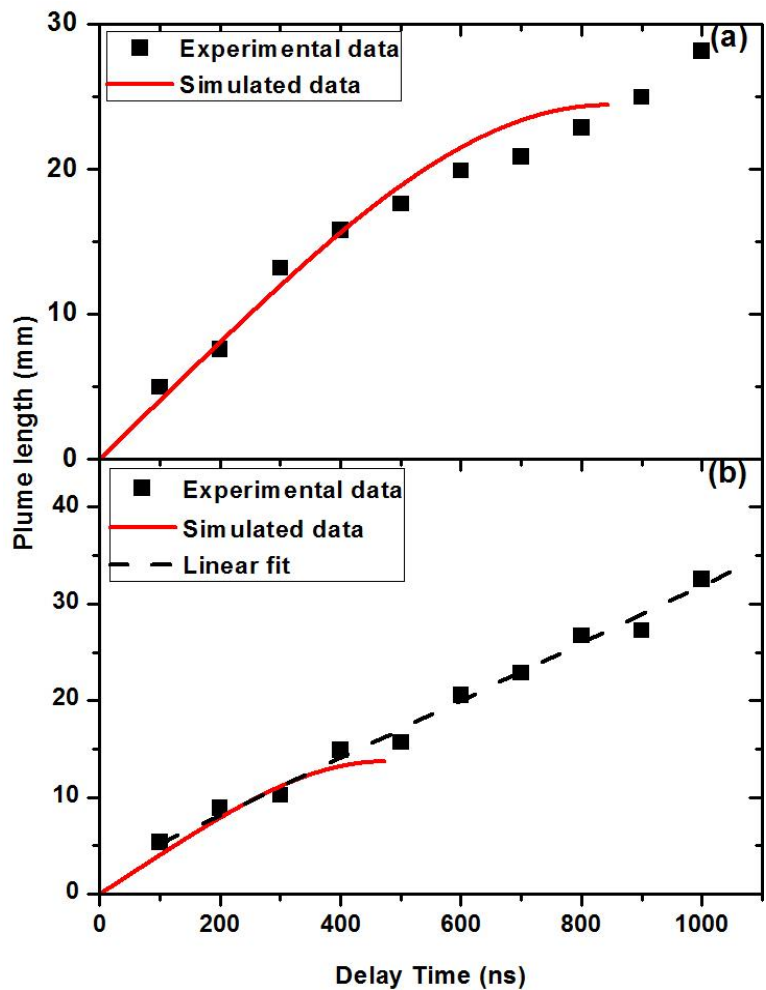

**Fig. 9.** Plume length as a function of time delay in the presence of (a) 0.21 T and (b) 0.5 T magnetic fields. Dash and solid lines represent the linear fit of experimental data and simulated data respectively.

Interestingly, increase of the magnetic field strength from 0.13 to 0.5 T does not bring significant changes in the case of carbon plasma plume. However, the slight decrease in the separation of Y-shaped bifurcation is observed with increase of magnetic field as shown in **Fig. 7b**. Similarly in case of time resolved data, both X and Y-images show nearly similar behaviour in 0.13 and 0.5 T magnetic field (**Fig. 5b** & **8b**) except little increase in the emission intensities at 0.5 T magnetic field in comparison to low magnetic field. It seems that the highest considered magnetic field is not sufficient to make any remarkable change in comparison to low magnetic field.

## 4. Conclusion

Time-resolved simultaneous captured two-directional images in the absence and presence of different magnetic fields varying from 0.13 to 0.5 T are analysed for the comparative study of interaction of magnetic field with carbon and copper plasma plume. The dynamical features, initiation and dismissal of diamagnetism of plume and subsequent structure formation are thoroughly discussed in these two cases. In contrast to the usual elliptical shaped copper plasma plume in the absence of a magnetic field, the carbon plasma plume splitted into slow and fast components which are attributed as neutrals and ion dominated regions respectively. At low magnetic field (0.13 T), deceleration of plume, formation of diamagnetic cavity due azimuthally drift of electrons (diamagnetic current) and its evolution and three-dimensional structure is clearly demonstrated in case of copper. On the other hand, diffusion of the magnetic field into the plasma plume followed by generation of electric field gradient and finally $E \times B$ is responsible for sharp conical structure and linear expansion of copper plume at 0.5 T magnetic field. The present observations suggest that carbon plasma did not show the diamagnetic behaviour at the considered magnetic field because of the weak restoring force required for the generation of radial electric field and hence the diamagnetic current. The Y-shaped structure of the carbon plasma plume is attributed as the polarization of charge dominated region in the diffused magnetic field whereas neutral component remains unaffected. Dependence of size and shape of the diamagnetic cavity on the field strength is clearly observed in copper plasma whereas increase of magnetic field from 0.13 to 0.5 T did not bring any significant change in structure of carbon plasma. Further the simulated dynamics of the plume in the magnetic field based on the modified energy balanced equation for the elliptic cylindrical plume geometry shows the close agreement with the present observations.